\documentclass[twocolumn,english,prx,showpacs,aps]{revtex4-1}

\usepackage[T1]{fontenc}
\usepackage[latin9]{inputenc}
\usepackage{amsmath}
\usepackage{amssymb}
\usepackage{graphicx}
\usepackage{esint}
\usepackage{pdfpages}
\usepackage{babel}

\makeatletter
 
 \@ifundefined{textcolor}{}
 {%
   \definecolor{BLACK}{gray}{0}
   \definecolor{WHITE}{gray}{1}
   \definecolor{RED}{rgb}{1,0,0}
   \definecolor{GREEN}{rgb}{0,1,0}
   \definecolor{BLUE}{rgb}{0,0,1}
   \definecolor{CYAN}{cmyk}{1,0,0,0}
   \definecolor{MAGENTA}{cmyk}{0,1,0,0}
   \definecolor{YELLOW}{cmyk}{0,0,1,0}
 }

\@ifundefined{textcolor}{}{%
 \definecolor{BLACK}{gray}{0}
 \definecolor{WHITE}{gray}{1}
 \definecolor{RED}{rgb}{1,0,0}
 \definecolor{GREEN}{rgb}{0,1,0}
 \definecolor{BLUE}{rgb}{0,0,1}
 \definecolor{CYAN}{cmyk}{1,0,0,0}
 \definecolor{MAGENTA}{cmyk}{0,1,0,0}
 \definecolor{YELLOW}{cmyk}{0,0,1,0}
 }

\@ifundefined{definecolor}{\@ifundefined{definecolor}
 {\@ifundefined{definecolor}
 {\@ifundefined{definecolor}
 {\@ifundefined{definecolor}
 {\@ifundefined{definecolor}
 {\usepackage{color}}{}
}{}
}{}
}{}
}{}
}{}
\makeatother 

\begin{document}

\title{Proposal for a motional-state Bell inequality test with ultracold atoms}

\author{R.~J.~Lewis-Swan}
\affiliation{The University of Queensland, School of Mathematics and Physics, Brisbane, Queensland 4072, Australia.}
\author{K.~V.~Kheruntsyan}
\affiliation{The University of Queensland, School of Mathematics and Physics, Brisbane, Queensland 4072, Australia.}
\date{\today }
\begin{abstract}
We propose and theoretically simulate an experiment for demonstrating a motional-state Bell 
inequality violation for pairs of momentum-entangled atoms produced in Bose-Einstein condensate collisions. The proposal is based on realizing an atom-optics analog of the Rarity-Tapster 
optical scheme: it uses laser-induced Bragg pulses to  
implement two-particle interferometry on the underlying Bell-state 
for two pairs of atomic scattering modes with equal but opposite momenta. The 
collision dynamics and the sequence of Bragg pulses are simulated using the stochastic Bogoliubov approach in the positive-$P$ representation. We 
predict values of the Clauser-Horne-Shimony-Holt (CHSH) parameter up to $S\simeq2.5$ for experimentally realistic parameter regimes, showing a strong 
violation of the CSHS-Bell inequality bounded classically by $S\leq2$.
\end{abstract}


\maketitle

\section{Introduction}

Bell inequalities \cite{Bell:64,*Bell:book} have arguably been regarded as ``the most profound discovery in science'' \cite{Stapp:75}. They provide a fundamental distinction between local hidden-variable (LHV) descriptions of physical reality and the description 
based on quantum mechanics wherein the concept of nonlocal entanglement is a fundamental ingredient. Violations of Bell inequalities, which reject all LHV theories and attest for the validity 
of quantum mechanics, have been demonstrated in numerous experiments with massless photons \cite{Aspect:82a,*Aspect:82b,Ou-Mandel-88,Rarity-Tapster:90,Zeilinger:98}, but in only a handful of experiments involving massive 
particles \cite{rowe2001experimental,sakai2006spin}. In addition, all massive particle experiments have so far been restricted to exploiting entanglement between internal (spin) degrees of freedom, but never between external (motional) degrees of freedom such as translational momentum. 
Here, we propose and simulate a 
matter-wave experiment which, for the first time, can demonstrate a Bell inequality violation for pairs of momentum-entangled ultracold atoms produced in a collision \cite{Perrin:2007,Krachmalnikoff:2010,Jaskula-10,Cauchy-Schwarz} of two Bose-Einstein condensates (BECs). In such a motional-state 
Bell inequality test,  particle masses become directly relevant, thus enabling extensions of fundamental tests of quantum mechanics into regimes which may involve couplings to gravitational fields and hence find connections to theories of gravitational decoherence \cite{Penrose:96}. 
This is important in view of future possible tests of quantum mechanics or its modifications (which currently go beyond established theories) in an attempt to resolve the current incompatibility of quantum mechanics and the theory of gravity.

The original Bell inequality was formulated by John Bell \cite{Bell:64,*Bell:book} in response to Einstein, Podolsky, and Rosen's (EPR) argument \cite{EPR:35,*Bohm:52} that, under the premises of local realism, quantum mechanics appears to be incomplete and hence must be supplemented by hidden variables in order to explain the `spooky-action-at-a-distance' due to entanglement between space-like separated particles. 
The first conclusive experimental demonstrations of Bell inequality violations with photons were reported in the early 1980s through to 1990s \cite{Aspect:82b,Ou-Mandel-88,Rarity-Tapster:90,Zeilinger:98} and used sources of pair-correlated photons, such as from a radiative cascade or parametric down-conversion. It took almost another two decades before the first massive-particle 
Bell violations emerged, utilizing pairs of trapped ions \cite{rowe2001experimental} or proton pairs from the radiative decay of metastable $^{2}$He \cite{sakai2006spin}. These experiments all relied on entanglement between the internal degrees of freedom---either the photon polarizations or the particle spins, with the notable exception of the Rarity-Tapster experiment \cite{Rarity-Tapster:90} which explored entanglement between photons momenta (see also \cite{Boyd:2004}).

In recent years, there has been an increasing number of experiments, particularly in the field of ultracold atoms \cite{Oberthaler:2008,Riedel-interferometer,Hannover-twins} 
and opto-mechanics \cite{Diamonds}, generating and quantifying various forms of massive-particle entanglement \cite{Polzik:2001,*Kuzmich:2006,Blatt-14:2011}. 
However, these should be distinguished from experiments designed to rule out LHV theories via a Bell inequality violation---the most stringent test of quantum mechanics. 
Ultracold atoms, nevertheless, provide a promising platform for extending these experiments towards Bell inequality tests \cite{Gneiting:2008,mullin2008pra,*mullin2009epj}, due to their high degree of isolation 
from the environment and the existing high degree of control over system parameters, including the internal and external degrees of freedom.

\begin{figure}[tbp]
\includegraphics[width=8.2cm]{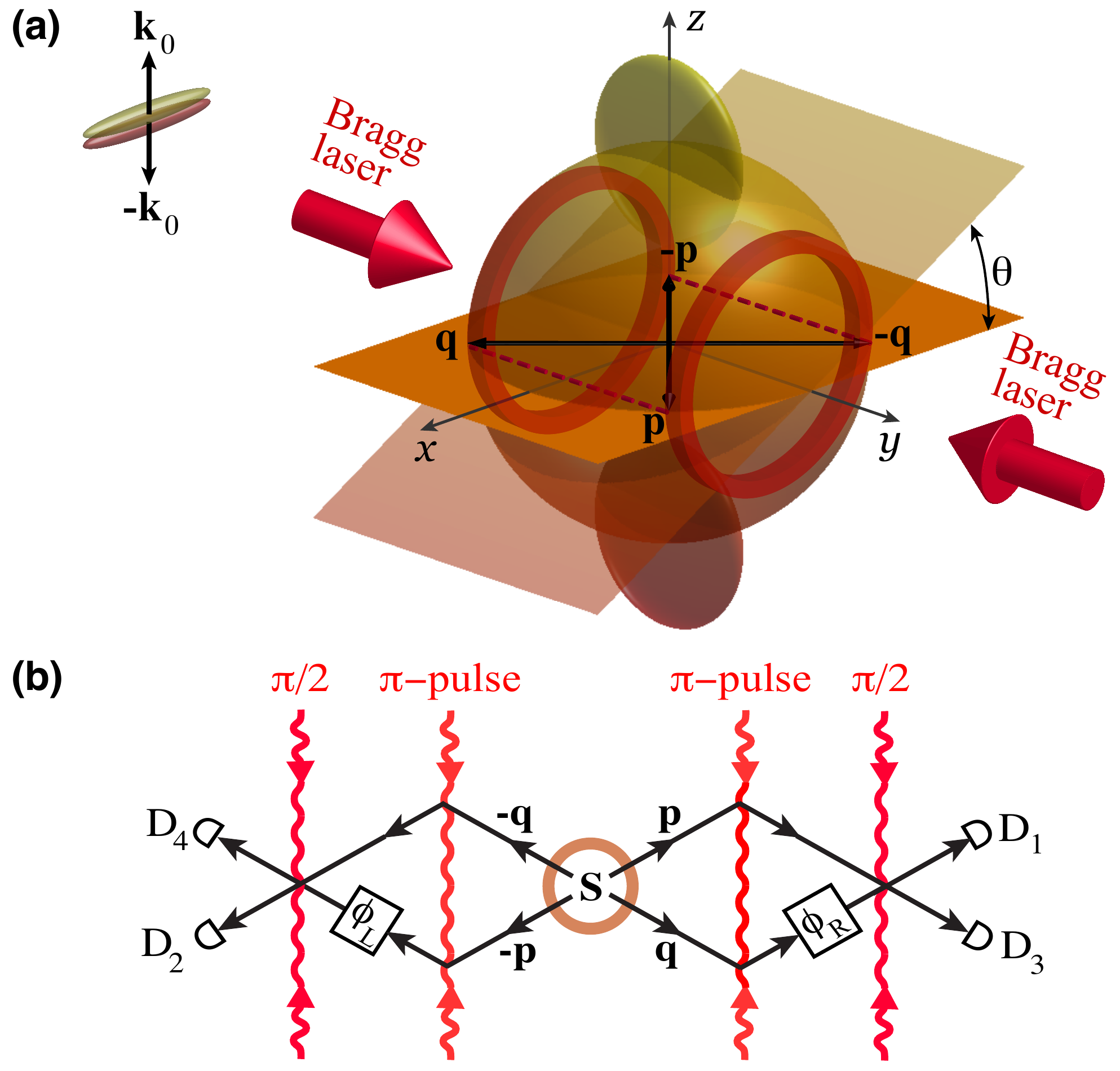}\caption{(Color online) Schematic diagram of the collision geometry and the proposed adaptation of the Rarity-Tapster scheme. (a) The two condensates in position space, 
counter-propagating along the $z$ axis with mean momenta $\pm\mathbf{k}_0$, are shown in the 
left, upper corner; the same condensates in momentum space (or after a time-of-flight expansion) have a pancake shape and are shown on the north and south poles of the spherical halo of scattered atoms. The counter-propagating (along $y$) Bragg lasers are tuned to couple and transfer the population between two pairs of momentum modes, such as the pair ($\mathbf{p,q}$) and ($\mathbf{-q,-p}$), indicated on the equatorial plane of the scattering halo. A similar quartet of modes (not shown for clarity), coupled by the same Bragg lasers, can be identified on any other plane obtained by rotating the equatorial plane by an angle $\theta$ around the $y$ axis; together, all these quartets of modes form two opposing rings shown in red. 
(b) The Rarity-Tapster scheme for implementing the $\pi$ and $\pi/2$ Bragg pulses on pairs of momentum modes emanating from the source (S) and the arrangement of two independent relative phase setting $\phi_L$ and $\phi_R$ (respectively, between $\mathbf{p}$ and $\mathbf{q}$, and between $\mathbf{-p}$ and $\mathbf{-q}$) imposed in the left and the right arms of the setup. After being mixed by the final $\pi/2$ pulse, the output modes are detected by four atom detectors $D_i$ ($i=1,2,3,4$) and different coincidence counts $C_{ij}$ are measured for calculating the CHSH-Bell parameter $S$.}
\label{fig:scheme}
\end{figure}

Our proposal for a motional-state Bell inequality uses pair-correlated atoms from colliding Bose-Einstein condensates and in this respect represents an ultimate successor to recent experiments demonstrating 
sub-Poissonian relative atom number statistics, violation of the classical Cauchy-Schwartz inequality \cite{Jaskula-10,Cauchy-Schwarz}, atomic Hong-Ou-Mandel effect \cite{Lewis-Swan:2014,AtomicHOM}, and a recent 
theoretical proposal for demonstrating the EPR paradox \cite{Zeilinger-EPR:2012} using the same collision process. A closely related process of dissociation of diatomic molecules has been recently proposed in 
Ref. \cite{Gneiting:2008} for demonstrating a Bell violation based on energy-time entanglement; the same process of molecular dissociation was previously discussed in Ref. \cite{Kheruntsyan:05} in the context of the EPR 
paradox for atomic quadrature measurements.

\section{Proposed atomic Rarity-Tapster setup}

The schematic diagram of the proposed experiment is shown in Fig.~\ref{fig:scheme}. A highly elongated (along the $x$ axis) BEC is
initially split into two counterpropagating halves with momenta
$\pm\mathbf{k}_{0}$ along $z$ in the center-of-mass frame \cite{Perrin:2007,Krachmalnikoff:2010}. Constituent
atoms of the condensate undergo binary elastic $s$-wave scattering and
populate a nearly spherical scattering halo (of radius $k_{r}\simeq
0.95|\mathbf{k}_{0}|$) of pair-correlated atoms \cite{Krachmalnikoff:2010} via the process of spontaneous four-wave mixing. Previous experiments and theory \cite{Perrin:2007,Perrin-08,Krachmalnikoff:2010,Jaskula-10,Cauchy-Schwarz} have
shown the existence of strong atom-atom correlation between pairs of diametrically opposite momentum modes, such as ($\mathbf{p,-p}$) and ($\mathbf{q,-q}$) (shown in Fig.~\ref{fig:scheme} on the equatorial plane of the scattering halo), similar to the correlation
between twin-photons in parametric down-conversion \cite{Ou-Mandel-88,Rarity-Tapster:90,Zeilinger:98}. After the end of the collision, we apply two separate Bragg pulses ($\pi$ and $\pi/2$) tuned to couple uncorrelated atoms from each 
respective pair, namely ($\mathbf{p,q}$) and ($\mathbf{-p,-q}$). The Bragg pulses replicate the atom optics analogs of a mirror and a beam splitter [see Fig. \ref{fig:scheme} (b)], thus realising the two interferometer arms of the Rarity-Tapster 
optical setup \cite{Rarity-Tapster:90} (see also Ref. \cite{Demler-Aspect:2011} which proposes the same scheme for implementing phase-sensitive measurements with ultracold atoms). A variable phase shift is additionally applied before the beam-splitter ($\pi/2$) pulse to the two lower arms of the interferometer, corresponding to a
relative phase shift of $\phi_L$ between $-\mathbf{p}$ and $-\mathbf{q}$, and $\phi_R=\phi_L+\phi$ between $\mathbf{q}$ and $\mathbf{p}$. This replicates the polarizer angle setting or relative phase settings in the optical Bell tests 
of Refs. \cite{Aspect:82b,Rarity-Tapster:90}, and can be realized by means
of introducing a relative phase $\phi_L$ between the two counterpropagating Bragg lasers that realize the $\pi$ pulse, combined with an additional relative phase shift $\phi$ between the left and the right arms of the interferometer, implemented by, e.g., the well-established technique of optical phase imprinting \cite{Sengstock:99,*Phillips:2000}.

In the low gain regime of atomic four-wave mixing (see below), this process approximately realizes a prototypical
Bell state of the form 
\begin{equation} 
|\Psi\rangle =\frac{1}{\sqrt{2}} (|1_{\mathbf{p}},1_{-\mathbf{p}}\rangle +|1_{\mathbf{q}},1_{-\mathbf{q}}\rangle ), 
\label{eq:Bell-state}
\end{equation} 
which corresponds to a pair of atoms in a quantum superposition of belonging to either the momentum modes $\mathbf{p}$ and $\mathbf{-p}$, or $\mathbf{q}$ and $\mathbf{-q}$. By measuring appropriate second-order correlation functions 
using atom-atom coincidences between certain pairs of atom detectors $D_i$ ($i=1,2,3,4$), for a chosen set of applied phases $\phi_L$ and $\phi_R$, one can construct (see below) the CHSH-Bell parameter $S$ for the Clauser-Horne-Shimony-Holt (CHSH) 
version of the Bell inequality \cite{CHSH:69,Aspect:82b}. The choice of phase settings $\phi_R$ and $\phi_L$ gives rise to non-locality in the vein of the original EPR paradox as atom-atom coincidences are intrinsically dependent on both 
phase settings, analogous to choosing polarization directions in archetypal optics experiments \cite{Aspect:82a,Aspect:82b}. 
Indeed, the Rarity-Tapster interferometric scheme can be mapped to a spin-$1/2$ or polarization-entangled system \cite{Aspect:82a}, 
wherein choosing the phases $\phi_L$ and $\phi_R$ directly controls the polarization basis in which each measurement is made.

Apart from coupling two pairs of momentum modes, ($\mathbf{p,q}$) and ($\mathbf{-q,-p}$), shown on the equatorial plane of Fig.~\ref{fig:scheme} (a), the Bragg pulses couple many other pairs of scattering modes that have the same wave-vector 
difference of $2k_r \!\equiv \!|\mathbf{p}-\mathbf{q}|\!=\!|(\mathbf{-p})-(\mathbf{-q})|$. Quartets of such modes, forming independent Bell states, can be identified on any 
other plane obtained from the equatorial plane by rotating it by an angle $\theta$ around the $y$ axis. Atom-atom coincidences between these modes can therefore be used as independent measurements for evaluating the respective 
CHSH-Bell parameter $S$. Averaging over many coincidence counts obtained in this way on a single scattering halo (in addition to averaging over many experimental runs) can be used to increase the signal-to-noise ratio and ultimately help 
the acquisition of a statistically significant result for $S$.

\section{Simple toy model}

Before presenting the results of our simulations, we make a brief diversion to discuss an important difference between the ideal prototype Bell state of the form of Eq. (\ref{eq:Bell-state}) 
and that which corresponds to the output of the simplest model of four-mode optical parametric down-conversion, to which our system can be reduced in its most rudimentary approximation (see Refs. \cite{Perrin-08,Ogren-09,*Savage:06} and Appendix \ref{AppA}). The 
Hamiltonian describing this process \cite{Walls:08,ReidWalls-86} can be written as $\hat{H} = \hbar{g}(\hat{a}_{1}^{\dagger}\hat{a}_{2}^{\dagger} +
\hat{a}_{3}^{\dagger}\hat{a}_{4}^{\dagger} + h.c.)$, where $g>0$ is a gain coefficient, related in our context 
to the density $\rho_0$ of the initial source condensate (assumed uniform) and the $s$-wave interaction strength $U=4\pi \hbar^2a/m$  through $g=U\rho_0/\hbar$ \cite{Perrin-08,Ogren-09}, where $a$ is the $s$-wave scattering length. The output state of this
model (for an initital vacuum state for all four modes) in the Schr{\"o}dinger picture can be written in terms of an expansion in the Fock-state basis as \cite{Braunstein-review,Lewis-Swan:2014} 
\begin{equation} 
|\Psi\rangle = (1-\alpha^2)\sum_{k,m=0}^{\infty} \alpha^{(k+m)} |k\rangle_1|k\rangle_2 |m\rangle_3|m\rangle_4, 
\label{eq:Psi-full} 
\end{equation} 
where
$\alpha = \mathrm{tanh}(gt)$ and $t$ is the collision duration. In the
weak-gain regime, which corresponds to $\alpha\simeq gt$ and hence an average mode occupation in each of the four modes 
($\langle \hat{a}_{i}^{\dagger} \hat{a}_{i}\rangle \equiv n=\sinh^2(gt)$, $i=1,2,3,4$) of $n \simeq \alpha^2 =(gt)^2 \ll 1$, the
sum over Fock states can be truncated to lowest order in $\alpha$ to 
\begin{eqnarray} 
|\Psi\rangle & \propto & |0\rangle_1|0\rangle_2|0\rangle_3|0\rangle_4 \nonumber \\ 
& + & \alpha (|1\rangle_1|1\rangle_2|0\rangle_3|0\rangle_4 + |0\rangle_1|0\rangle_2|1\rangle_3|1\rangle_4).
\label{eqn:Psi} 
\end{eqnarray}

Taking into account the fact that the contribution from the pure vacuum state (the first term) does not affect the outcome of 
any correlation (coincidence) measurements (except for reducing the absolute data acquisition rate through multiple experimental realizations), we can further approximate this state by
$|\Psi\rangle \propto \alpha (|1\rangle_1|1\rangle_2|0\rangle_3|0\rangle_4 + |0\rangle_1|0\rangle_2|1\rangle_3|1\rangle_4)$. 
Equation (\ref{eq:Bell-state}) corresponds to this state in a shorthand notation.
Such a state can itself be mapped to the archetypical Bell 
state $|\Psi^{+}\rangle = \frac{1}{\sqrt{2}}(|+\rangle_{L}|-\rangle_{R} + |-\rangle_{L} |+\rangle_{R})$ 
in the polarization or spin-$1/2$ $\hat{S}_z$ basis, where the subscript ($L,R$) refers to the left and right arms of the interferometer and $+$ ($-$) refer 
to the upper (lower) paths, in terms of the diagram of Fig. \ref{fig:scheme} (b) of the main text.

This ideal Bell state gives a maximal value of $S=2\sqrt{2}$ (for a definition of the CHSH-Bell parameter $S$, see Sec. 
\ref{Main}) and hence a maximal Bell violation ($S>2$) by definition. However, in general, 
when using spontaneous parametric down-conversion as a suitable source of pair correlated particles, 
one must keep in mind the contribution from the higher-order Fock states (whose relative weight is very small for $n\ll 1$, 
implying that the contribution of events that produce, e.g., two or more photons in each of the correlated modes is extremely unlikely), 
leading to a breakdown of the mapping of the full state Eq. (\ref{eq:Psi-full}) to Eq. (\ref{eq:Bell-state}) and thus a 
reduction in $S$ from the maximum value of $2\sqrt{2}$ to
\begin{equation} 
S = 2\sqrt{2}\frac{1+n}{1+3n}.
\label{eqn:S_undep_pump} 
\end{equation} 
This expression corresponds, in fact, to the full output state, Eq. (\ref{eq:Psi-full}), without any truncation of higher-order Fock states, and hence is valid for arbitrary $n$; 
it follows (see Appendix \ref{AppA}) from the maximally valued anomalous moment $|m|^2\equiv |\langle \hat{a}_1\hat{a}_2\rangle|^2 =|\langle \hat{a}_3\hat{a}_4\rangle|^2 =n(n+1)$, 
which is the case for this simple parametric down-conversion model \cite{Ogren-09,*Savage:06}, where $n=\sinh^2(gt)$.

Equation (\ref{eqn:S_undep_pump}) is an insightful result from the simplest analytic treatment as it shows the scaling of
$S$ with the mode population: for $n\ll1$ we indeed obtain a nearly maximal Bell violation, $S\simeq 2\sqrt{2}$
while we find an upper bound of $n=n_{\mathrm{cr}}=(\sqrt{2}-1)/(3-\sqrt{2})\simeq0.26$ beyond which 
the violation is no longer observed as $S\leq 2$ for $n\geq n_{\mathrm{cr}}$. 
We thus conclude that, for a large Bell violation, it is necessary to work in the low gain, low mode occupation regime of $n\ll 1$, which has, however, a practical 
inconvenience of requiring a large number of repeated experimental runs for achieving statistically significant data acquisition rate.

\section{Stochastic Bogoliubov simulations: results and discussion}
\label{Main}

\begin{figure}[tbp]
\includegraphics[width=7.3cm]{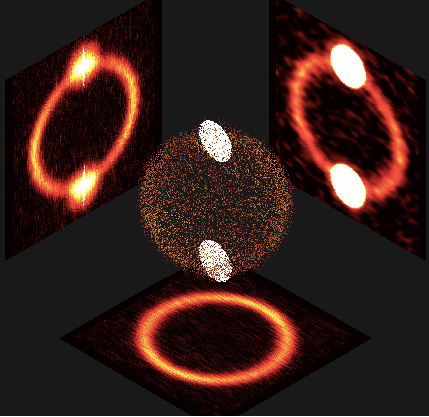}
\caption{(Color online) Illustration of typical results for the collisional halo in momentum space from the stochastic Bogoliubov approach in the positive-$P$ representation. Shown here are three orthogonal slices (cuts through the origin) of the 3D momentum distribution $n(\mathbf{k})$ at the end of the collision; the saturated (white) regions of the color map correspond to the high-density colliding condensates. The central figure is a discretized scatter plot of the 3D data (shown only for illustrational purposes and comparison with Fig. \ref{fig:scheme}), in which the dots (pixels) represent random samples of the average, but still fluctuating within the sampling error, density distribution binned into pixels whose color coding scales with the atom number in the bin (only four color grades were used for clarity). For quantitative details of the same data on the equatorial plane, see Fig. \ref{fig:Results}.}
\label{fig:halo} 
\end{figure}

\begin{figure*}[tbp]
\includegraphics[width=17.5cm]{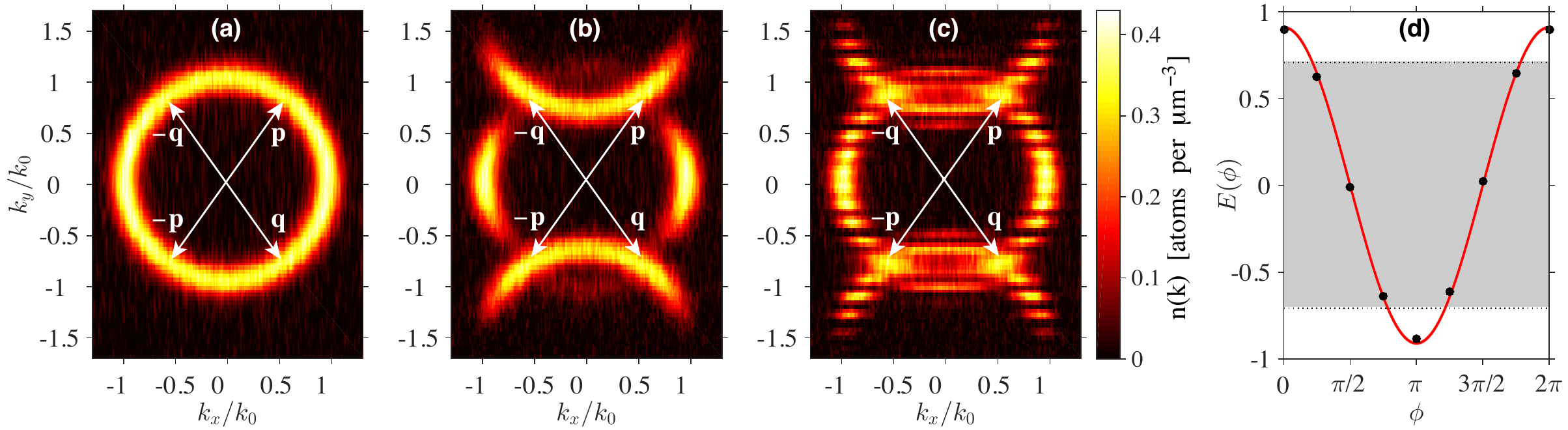}
\caption{
(Color online) Momentum distribution $n(\mathbf{k})$ of scattered atoms on the equatorial plane of the halo and the correlation coefficient $E$. Momentum distribution is shown (a) after the collision, at $t_1\!=\!65$ $\mu$s; (b) after the $\pi$ pulse chosen here to be a Gaussian, centered
at $t_2\!=\!79$ $\mu$s and having a duration (rms
width) of $\tau_{\pi}\!=\!3.5$ $\mu$s; and (c) after the final $\pi/2$ pulse,
centered at $t_{3}\!=\!139$ $\mu$s 
and having a duration of $\tau_{\pi/2}\!=\!3.5$ $\mu$s.
The momentum axes $k_{x,y}$ 
 are normalized to the collision momentum 
$k_{0}\!\equiv\!|\mathbf{k_{0}}|$ (in wave-number units), which in our
simulations is $k_{0}\!=\!4.7\!\times\!10^{6}$ m$^{-1}$. The plotted results
are for an initial BEC containing a total average number of $N\!=\!1.9\! \times \!10^4$ atoms
of metastable helium ($^{4}$He$^{*}$) prepared in a harmonic trap of
frequencies $(\omega_{x},\omega_{y},\omega_{z})/2\pi\!=\!(64,1150,1150)$ Hz
and colliding with the scattering length of $a\!=\!5.3$ nm; all these parameters are very close to those realized in recent experiments 
\cite{Krachmalnikoff:2010,Jaskula-10,Cauchy-Schwarz}. The optimal timing of the final Bragg pulse differs slightly for condensates with different $N$; in particular, $t_{3}$ 
ranged from $135.5$ to $139$ $\mu$s for the data in Fig.~\ref{fig:S} (see Appendix \ref{AppB}).
The data are averaged over $\sim\!30,000$ stochastic trajectories on a spatial lattice of $722\times192\times168$ points.
Panel (d) shows the correlation coefficient $E(\phi_L,\phi_R)$ as a function of $\phi\! \equiv\! \phi_L \!- \!\phi_R$, for the same detection bin sizes as in Fig.~\ref{fig:S}, blue circles. 
The data points are from numerical simulations (error bars of two standard deviations, representing sampling errors from $360$ 
stochastic runs, are within the marker size), including averaging over $\sim 370$ quartets of distinct detection volumes on the two opposing rings of the scattering halo shown in Fig. \ref{fig:scheme}, 
while the solid line is from the Gausssian-fit model, Eq.~(\ref{eqn:integrated_E}). A maximum amplitude of 
$E_{0}\!>\! 1/\sqrt{2}$ (outside the shaded region) corresponds to a correlation strength that can lead to a Bell inequality violation, 
given the underlying sinusoidal behavior. 
}
\label{fig:Results} 
\end{figure*}

To simulate the generation and detection of Bell states via the proposed scheme we use the stochastic Bogoliubov approach in the positive-$P$ representation \cite{Krachmalnikoff:2010,Deuar-11}, in which the scattered atoms are 
described by a small fluctuating component $\hat{\delta}(\mathbf{r},t)$ in the expansion of the full field operator $\hat{\Psi}(\mathbf{r},t)=\psi_0(\mathbf{r},t)+\hat{\delta}(\mathbf{r},t)$, where $\psi_0(\mathbf{r},t)$ is 
the mean field component describing the source condensate assumed to be in a coherent state of total average number $N$, initially in the ground state of the confining trap potential.
This approach has previously been used to accurately 
model a number of condensate collision experiments, including the measurement and characterisation of atom-atom correlations via sub-Poissonian relative number statistics \cite{Jaskula-10}, violation of the classical Cauchy-Schwarz 
inequality \cite{Cauchy-Schwarz}, and more recently in a theoretical proposal for demonstrating an atomic Hong-Ou-Mandel effect \cite{Lewis-Swan:2014}. The positive-$P$ representation has also been used in Ref. \cite{Drummond-Bell-PRA:2014} for direct probabilistic 
sampling of an idealised, polarization-entangled Bell state to show how a Bell inequality violation can be simulated using the respective phase-space distribution function. Complementary to Ref. \cite{Drummond-Bell-PRA:2014}, we do not assume any 
pre-existing Bell state in our analysis, but adopt an operational approach of calculating a set of pair-correlation functions $C_{ij}$ that define the CHSH-Bell parameter $S$, after real-time simulations of the collision dynamics and the application 
of Bragg pulses. (For the most recent formulation of the stochastic positive-$P$ equations that we simulate, including the application of the lattice potential imposed by the Bragg lasers, see the Methods section of Ref. \cite{Lewis-Swan:2014}.)

The CHSH-Bell parameter $S$ corresponding to our measurement protocol, performed for four pairs of phase settings, is defined as \cite{CHSH:69,Rarity-Tapster:90}
\begin{equation} 
S =|E(\phi_L,\phi_R) - E(\phi_L,\phi_R') +
E(\phi_L',\phi_R) + E(\phi_L',\phi_R')|, 
\label{eq:S}
\end{equation} 
where
\begin{equation}
E(\phi_{L},\phi_{R})\equiv \left.\frac{C_{14}+C_{23}-C_{12}-C_{34}}{C_{14}+C_{23}+C_{12}+C_{34}}\right|_{\phi_{L},\phi_{R}}.
\label{eq:E}
\end{equation}
Here, the correlation functions $C_{ij}$ are given by $C_{ij}\!=\! \langle \hat{N}_i
\hat{N}_j \rangle$, where the operator $\hat{N}_i(t) \!=\! \int_{\mathcal{V}(\mathbf{k}_i)}
d^3\mathbf{k} ~ \hat{n}(\mathbf{k},t)$ corresponds to the number of atoms detected in
a detection bin with dimensions $\Delta k_d$ ($d=x,y,z$) and volume $\mathcal{V}(\mathbf{k}_i)=\prod_d \Delta k_d$, centered around the targeted momenta $\mathbf{k}_i$ ($i \!=\! 1,2,3,4$);
the set of momenta $\{\mathbf{k}_1,\mathbf{k}_2,\mathbf{k}_3,\mathbf{k}_4 \}$ correspond, respectively, to  $\{\mathbf{p},\mathbf{-p},\mathbf{q},\mathbf{-q} \}$ used in the diagram of Fig.~\ref{fig:scheme}, while $\hat{n}(\mathbf{k},t)=\hat{a}^{\dagger}(\mathbf{k},t)
\hat{a}(\mathbf{k},t)$ is the momentum-space density, with $\hat{a}(\mathbf{k},t)$ being the Fourier component of the field operator $\hat{\delta}(\mathbf{r},t)$ describing the scattered atoms. The CHSH-Bell inequality states that any LHV theory satisfies an upper bound given by $S\leq 2$, irrespective of the phase settings $\phi_L$, $\phi_R$, $\phi_L'$, and $\phi_R'$.

The results of our numerical
simulations of the collision dynamics and ensuing Bragg pulses are shown in 
Figs.~\ref{fig:halo} and \ref{fig:Results}. Figure \ref{fig:halo} illustrates the momentum space density distribution of the collisional halo, while Fig. \ref{fig:Results} focuses on the quantitative results on the equatorial plane, for the following: (a) at the end of the collision; (b) after the application of the $\pi$ pulse; and (c) after the $\pi/2$ pulse.
The upper and lower semicircles in (b) correspond to Bragg-kicked
populations between the targeted momenta around $\mathbf{p}$ and $\mathbf{q}$, 
and between $\mathbf{-q}$ and $\mathbf{-p}$, while (c) shows the final
distribution after mixing. The density modulation in (c) (in parts of the halo lying 
outside the vicinity of the targeted momentum modes, where 
the transfer of population during the $\pi$ pulse is not $100$\% 
efficient) is simply the result of interference between the residual and transferred
atomic populations upon their recombination 
on the beam splitter \cite{Lewis-Swan:2014}.

We next use the stochastic Bogoliubov simulations to calculate the atom-atom correlations $C_{ij}$, for the optimal choice of phase 
angles $\phi_L \!=\! 0$, $\phi_L'\!= \!\pi/2$, $\phi_R\! =\! \pi/4$, and $\phi_R' \!= \!3\pi/4$ \cite{Rarity-Tapster:90}. 
The dependence of the resulting correlation coefficient $E$ on the relative phase $\phi\!\equiv \! \phi_L \!- \!\phi_R$ is shown in Fig.~\ref{fig:Results} (d); it displays a sinusoidal dependence $E_{0}\cos\phi$ which can also be predicted from a simple Gaussian-fit analytic model (see Appendix \ref{AppB}):
\begin{equation}
E\left( \phi_L,\phi_R \right) = \frac{ h\prod_d\alpha_d }{ h \prod_d \alpha_d + 2\prod_d (\lambda_d)^2 } \mathrm{cos}\left( \phi_L - \phi_R \right).
\label{eqn:integrated_E} 
\end{equation}
In this model, $C_{ij}$ is expressed in terms of the density-density correlation function $G^{(2)}(\mathbf{k},\mathbf{k}',t_1)\!=\!\langle \hat{a}^{\dagger}(\mathbf{k},t_1)\hat{a}^{\dagger}(\mathbf{k}',t_1)
\hat{a}(\mathbf{k}',t_1)
\hat{a}(\mathbf{k},t_1) \rangle$ after the collision  as $C_{ij}\!\!=\!\!\int_{\mathcal{V}(\mathbf{k}_i)}\!d^3\mathbf{k}\int_{\mathcal{V}(\mathbf{k}_j)}\!d^3\mathbf{k}' G^{(2)}(\mathbf{k},\mathbf{k}',t_1)$, and we use the fact that $G^{(2)}(\mathbf{k},\mathbf{k}',t_1)$ itself is typically well approximated \cite{Perrin:2007,Trippenbach:2008,Cauchy-Schwarz} by a Gaussian function of the form $G^{(2)}(\mathbf{k},\mathbf{k}',t_1)\!\!=\!\!\bar{n}^2 (1 + h \prod_d \mathrm{exp}[-(k_d+k'_d)^2/2
\sigma_{d}^2])$, where we have assumed that the density of scattered atoms is approximately constant over the integration volume and is given by $\bar{n}$. Thus, in Eq. (\ref{eqn:integrated_E}), $h$ is 
the height (above the background level of $\bar{n}^2$) of the pair correlation $G^{(2)}(\mathbf{k},\mathbf{k}',t_1)$, $\sigma_{d}$ is the rms width, $\lambda_d \equiv \Delta k_d/2\sigma_d$ is the relative bin size, and $\alpha_d  \equiv  ( e^{-2\lambda^2_d} - 1 ) + \sqrt{2\pi}\lambda_d ~ \mathrm{erf} \left( \sqrt{2}\lambda_d \right)$. The particular form of $E$ in Eq.~(\ref{eqn:integrated_E}) is obtained from this model by assuming the subsequent `mirror' and `beam-splitter' mix the coupled modes exactly. The visibility of the correlation coefficient $E$ bounds 
the maximum attainable violation of the CHSH-Bell inequality for a specific set of phase settings, with a lower-limit of $E_0 = 1/\sqrt{2}$ required for $S > 2$, and a maximum value of $E_0 = 1$ corresponding to $S=2\sqrt{2}$.

\begin{figure}[tbp]
\includegraphics[width=8.3cm]{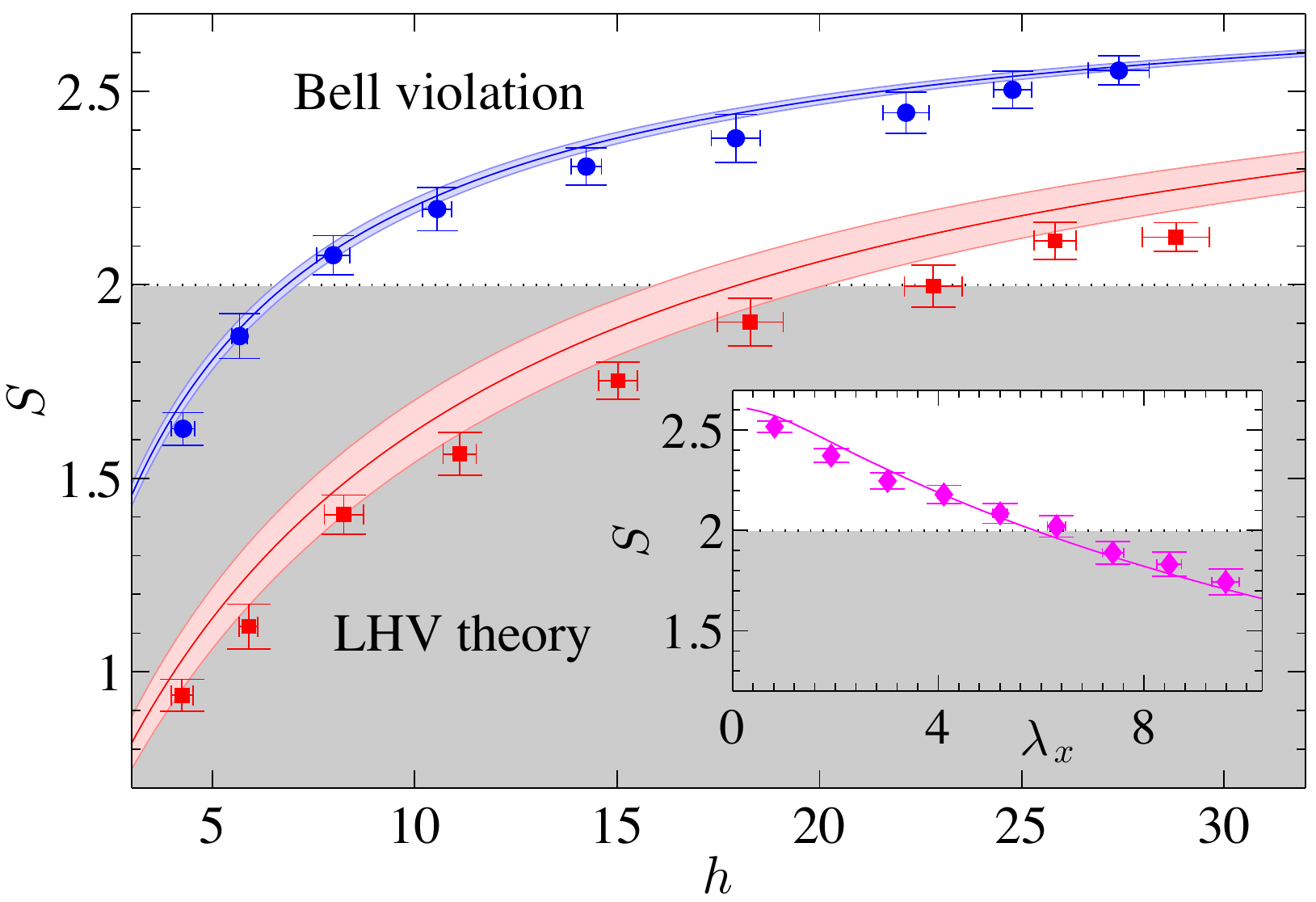}
\caption{CHSH-Bell parameter $S$ as a function of the correlation strength $h$ (see text); the value of $h$ can be controlled by varying the total average atom number $N$ in the initial BEC. For the data points shown here, $N$ was varied between $1.9 \!\times\! 10^4 $ (largest $h$) and $7.4\! \times\! 10^4$ (smallest $h$).
The two sets of data correspond to two different detection bin sizes: $(\Delta k_x, \Delta k_y, \Delta k_z)\! =\! (0.052, 0.53, 0.47)$~$\mu$m$^{-1}$ circles (blue) and  $(0.12, 1.24, 1.10)$~$\mu$m$^{-1}$ squares (red). The vertical error bars on data points indicate the stochastic sampling errors \cite{cpu_S}; the horizontal error bars are the sampling errors on the value of $h$. The results are compared to the analytic predictions (solid lines) of Eq. (\ref{eqn:integrated_CHSH}); 
uncertainty (shaded regions) is due to the uncertainty in determining $\sigma_d$. The inset shows the explicit dependence of $S$ on $\Delta k_x$ (in units of $2\sigma_x\!=\!0.068$~$\mu$m$^{-1}$), for fixed $(\Delta k_y,\Delta k_z)\!=\!(0.77\sigma_y,0.89\sigma_z)\!=\!(0.53, 0.47)$~$\mu$m$^{-1}$ and $N\!=\!1.9\times 10^4$ ($h\simeq 27$). 
For a typical time-of-flight expansion time of $t_{\mathrm{\exp}}\sim300$ ms, which maps the atomic momentum distribution into position space density distribution, and which is when the atoms are experimentally detected, these detection bin sizes convert 
to position space distances of $(\Delta x, \Delta y, \Delta z)\simeq(0.32, 2.5, 2.2)$ mm (where we have taken $\lambda_x=1$ for definitiveness), which are several times larger than the three orthogonal resolutions of multichannel plate detectors 
used in $^4$He$^*$ experiments \cite{Cauchy-Schwarz,Dall:13}.
} 
\label{fig:S} 
\end{figure}

The results of calculations of the CHSH-Bell parameter $S$ are shown in Fig.~\ref{fig:S}, where we explore its dependence on the strength of atom-atom correlations and the detection bin size. The dependence on the correlation strength, for a fixed collision velocity and trap frequencies, reflects essentially the dependence on the peak density of the initial BEC, which itself depends on the total average number of atoms loaded in the trap \cite{Perrin-08}.
The results of stochastic simulations in Fig.~\ref{fig:S} (b) are plotted alongside the predictions of the Gaussian-fit analytic model, which from Eq. (\ref{eqn:integrated_E}) gives
\begin{equation}
S = 2\sqrt{2}\;\frac{ h\prod_d
\alpha_d }{h \prod_d \alpha_d + 2\prod_d (\lambda_d)^2 }.
\label{eqn:integrated_CHSH} 
\end{equation}
As we see, the analytic prediction agrees reasonably  well with the numerical results; both show that strong Bell violations are favoured for: (i) smaller condensates, leading to lower mode population in the scattering halo and thus 
higher correlation strength, and (ii) smaller bin sizes, for which the strength of atom number correlations does not get diluted due to the finite detection resolution. The discrepancies between the numerical and analytic results are due to the fact that the analytic model assumes uniform halo density across the integration bin and perfect Bragg pulses, both in terms of the intended transfer efficiency and its insensitivity to the momentum offsets within the integration bin, whereas the numerical simulations are performed with realistic Bragg pulses acting on the actual inhomogeneous scattering halo. Nevertheless, an important conclusion that we reach here is that the Bell violation in our scheme can tolerate experimentally relevant imperfections that are often ignored in oversimplified models.

The general form of Eq.~(\ref{eqn:integrated_CHSH}) displays similar behaviour to that obtained in the simple model of four-mode parametric down-conversion, Eq. (\ref{eqn:S_undep_pump}). As previously, it gives a simple and insightful picture in terms of the dependence of the expected value of $S$ on just a few parameters at the end of the collision---the correlation widths, the correlation height and the detection bin size. As we see from the comparison of the predictions of Eq. (\ref{eqn:integrated_CHSH}) to the actual numerical results in Fig. \ref{fig:S}, the agreement is remarkable for such a simple analytic result. The scaling with the halo mode occupation, as that in Eq. (\ref{eqn:S_undep_pump}), is no longer explicit, but it now emerges most simply through the detection bin size, wherein a smaller bin size gives a smaller average number of detected atoms and hence larger values of $S$ as seen in the inset of Fig. \ref{fig:S}. Similarly, such a scaling emerges through the height of the correlation $h$:  the correlation is typically stronger for four-wave mixing regimes that produce a collisional halo of smaller density or smaller bin occupation (for a fixed bin size), leading to larger values of $S$. In the four-mode down-conversion model, where  the relevant normalized pair-correlation function is given by $g^{(2)}_{12}\!=\!g^{(2)}_{34}\!=\!2+1/n$ \cite{Ogren-09,*Savage:06} and therefore  $h\!=\!1\!+\!1/n$, this corresponds to $h\gg 1$ which is again the regime of low mode occupation $n\ll 1$ as we discussed previously.

We further emphasise that the general applicability of our Gaussian-fit analytic model and, in particular, the relatively simple result of Eq. (\ref{eqn:integrated_CHSH}) are not limited to condensate collision experiments. Rather, these results can be applied to any other ultracold atom experiment---a candidate for a Bell test---as long is it produces two pair-correlated `scattering' modes that can be approximated by Gaussian correlation functions and subsequently subjected to `mirror' and `beam-splitter' pulses to realize an atomic Rarity-Tapster interferometer.

~

\section{Conclusions}

In summary, we have shown that condensate collisions are a promising platform for testing motional-state Bell inequalities with massive particles. 
We predict a CHSH-Bell inequality violation ($S>2$) for a range of parameters well within currently accessible experimental regimes.

Our numerical simulations take into account a range of physically important processes beyond the common analysis of oversimplified toy models. Importantly this includes: (i) the multimode nature of the colliding Bose-Einstein condensates and subsequent scattering halo; 
(ii) the spatial expansion and separation of the source condensates during the collision and hence during the pair production process (for comparison, the `pump mode' in the optical down-conversion case remains practically unchanged in the required weak-gain regime); 
and (iii) the fact that the atomic `mirror' and `beam-splitter' Bragg pulses act, in fact, as momentum kicks (translations) rather than as actual (optical) reflections. By modeling the real-time application of the Bragg pulses, without 
assuming ideal $\pi$ and $\pi/2$ pulses (100\% and 50\% transfer, respectively), we implicitly allow for small amounts of losses (hence decoherence) into  higher-order Bragg scattering modes. We also take into account the nontrivial effects of phase dispersion, 
absent in photonic experiments, by optimizing the timing and application of the Bragg pulses in the interferometer. Remarkably, many of these effects can also be captured via the semi-analytic Gaussian-fit model of Eqs. (\ref{eqn:integrated_E}) and (\ref{eqn:integrated_CHSH}), 
which is found to be both qualitatively and quantitatively rather accurate.

Such detailed quantitative analysis is important for a theoretical proposal to be relevant to possible experimental demonstrations of a Bell inequality violation. This is further supported by our analysis in terms of finite detector resolution 
and the utilization of multiple quartets of bins in our calculations: increasing the rate of data acquisition is crucial for experiments with ultracold atoms which typically have relatively slow duty cycles of the order of half a minute (for comparison, 
the repetition rates of a pump laser in modern optical parametric down-conversion experiments can reach tens of MHz). 

A laboratory demonstration of such a violation would be a major advance in experimental quantum physics as it would lead to a better understanding of massive particle entanglement involving motional states. Apart from extending foundational tests of quantum mechanics into new regimes, such experiments
can potentially lead to an opening of a new experimental agenda, such as testing the theories of decoherence due to coupling to 
gravitational fields \cite{Penrose:96} and answering questions that are relevant to the understanding of the interplay between quantum theory and gravity and their possible unification.

\begin{acknowledgments}
The authors acknowledge stimulating discussions with A. Aspect and C. Westbrook, and the developers of XMDS2 software \cite{xmds2} used in the simulations. R. J. L-S also particularly acknowledges fruitful discussions with M. E. Lewis.
K. V. K acknowledges support by the Australian Research Council Future Fellowship Grant No. FT100100285. 
\end{acknowledgments}

\appendix

\section{Undepleted pump approximation and relation to the model of spontaneous parametric down-conversion. \label{sec:undep_pump}}  
\label{AppA}

The simplest analytic treatment of the scheme can be made by treating
the initially split condensate in the undepleted pump approximation \cite{Perrin-08},
corresponding to short collision durations such that the number of
scattered atoms is only a small fraction of the source condensate
(generally less than $10\%$). Treating the $\pi$ and $\pi/2$ Bragg pulses as perfect mirrors 
and beam splitters (i.e., simple linear transformations) applied at $t_2$ and $t_4$, respectively (see main text for definitions) and then invoking Wick's theorem, the 
second-order correlation function between the relevant pairs of detectors (chosen for definiteness to be equal to $t_4 =t_3 + 4\tau_{\pi/2}$ in our simulations) can be written as
\begin{eqnarray} 
& & G^{(2)}(\mathbf{k}_1,\mathbf{k}_2,t_4) =  G^{(2)}(\mathbf{k}_3,\mathbf{k}_4,t_4) \nonumber \\ 
& = &n(\mathbf{k}_1,t_1)^2  + \frac{|m(\mathbf{k}_1,\mathbf{k}_2,t_1)|^2}{2} \left[ 1 -
\mathrm{cos}\left( \phi_B - \phi_A \right) \right],  \;\; \;\;\;\; \label{eqn:G2_12_toy} \\
& & G^{(2)}(\mathbf{k}_1,\mathbf{k}_4,t_4)  =  G^{(2)}(\mathbf{k}_2,\mathbf{k}_3,t_4) \nonumber \\
& = & n(\mathbf{k}_1,t_1)^2  + \frac{|m(\mathbf{k}_1,\mathbf{k}_2,t_1)|^2}{2} \left[ 1 +
\mathrm{cos}\left( \phi_B - \phi_A \right) \right], \;\; \;\;\;\; 
\end{eqnarray}
where $n(\mathbf{k},t_1) = \langle
\hat{a}^{\dagger}(\mathbf{k},t_1)\hat{a}(\mathbf{k},t_1) \rangle$ is the average
momentum-space density of scattered atoms after the collision at time $t_1$, which is equal for the targeted modes $\textbf{k}_1, \textbf{k}_2, \textbf{k}_3$ and $\textbf{k}_4$, and
$m(\mathbf{k},\mathbf{k}',t_1) = \langle
\hat{a}(\mathbf{k},t_1)\hat{a}(\mathbf{k}',t_1) \rangle$ is the average anomalous
moment. Choosing $\phi_A = 0$, $\phi_A'
= \pi/2$, $\phi_B = \pi/4$, and $\phi_B' = 3\pi/4$ to maximize the CHSH-Bell
parameter $S$ (defined as per the main text) we find the result 
\begin{equation} 
S = 2\sqrt{2}
\frac{|m(\mathbf{k}_1,\mathbf{k}_2,t_1)|^2}{2n(\mathbf{k}_1,t_1)^2 +
|m(\mathbf{k}_1,\mathbf{k}_2,t_1)|^2}. 
\end{equation} For a maximal
violation, with $S=2\sqrt{2}$, one requires the anomalous moment to satisfy
$|m(\mathbf{k}_1,\mathbf{k}_2,t_1)|^2 \gg n(\mathbf{k}_1,t_1)^2$,
corresponding to strong correlations between atoms scattered to
diametrically opposite momentum modes.

The anomalous moment is maximized
for the case of a homogeneous BEC in a finite box \cite{Perrin-08,Savage:06}, where the discrete mode counterpart of $m(\mathbf{k},-\mathbf{k})$ satisfies
$|m_{\mathbf{k},-\mathbf{k}}|^2 = n_{\mathbf{k}}(1+n_{\mathbf{k}})$ \cite{Savage:06}---just like in the simple four-mode model of parametric down-conversion discussed in the main text, thus giving the result of Eq. (\ref{eqn:S_undep_pump}), 
with $n = n_{\mathbf{k}_i}$ 
($i = 1,2,3,4$) being the average mode occupation of the scattering halo after the
collision, which are all equal in this approximation.

\section{Gaussian-fit analytic model of correlation functions}
\label{AppB}

Beyond the simple treatment of the previous section, we can develop a more sophisticated model of the CHSH-Bell parameter while also taking into account 
the finite detector resolution of experiments \cite{Perrin:2007}. We calculate integrated pair-correlation functions and the ensuing CHSH-Bell parameter by using a Gaussian-fit analytic model, 
similar to that used previously in Ref. \cite{Cauchy-Schwarz} to model a violation of the Cauchy-Schwarz inequality 
in condensate collisions. The underlying assumption of the model is that the second-order correlation function 
after the collision is well approximated by a Gaussian $G^{(2)}(\mathbf{k},\mathbf{k}',t_1) = n^2(1 + h\prod_d \mathrm{exp} [-(k_d + k'_d)^2/2\sigma^2_d])$ 
for $\mathbf{k} \simeq -\mathbf{k}'$ and $n = n(\mathbf{k}) = n(\mathbf{k}')$ is the density of scattered atoms. The correlation is then characterised by two parameters: 
the height, $h$, above the background level and the correlation width $\sigma_d$. 

To derive an expression for $S$ we first consider the form of the integrated pair-correlation functions after the application of the $\pi/2$ pulse,
\begin{eqnarray}
 C_{ij} \!=\! \langle \hat{N}_i \hat{N}_j \rangle \!=\! \int_{\mathcal{V}(\mathbf{k}_i)} \!d^3\mathbf{k} \int_{\mathcal{V}(\mathbf{k}_j)} \!d^3\mathbf{k}' G^{(2)}\left( \mathbf{k},\mathbf{k}',t_4 \right), \;\; \;\;\;\;  \label{eqn:NiNj}
\end{eqnarray}
where the integration bins are of dimension $\Delta k_d$ ($d=x,y,z$) and volume $\mathcal{V}(\mathbf{k}_i) = \prod_d \Delta k_d$ centered around the targeted momenta $\mathbf{k}_i$ ($i=1,2,3,4$). 
Without loss of generality we consider the form of the correlation $C_{12}$, with the remaining pair-correlation functions $C_{ij}$ being calculated in a similar manner. 
Treating the Bragg pulses as idealised mirrors and beam-splitters which act instantaneously, meaning we may set $t_2 = t_1$ and $t_4 = t_3$, we may write the generalised form of Eq. (\ref{eqn:G2_12_toy}) as
\begin{widetext}
\begin{eqnarray}
 G^{(2)}(\mathbf{k},\mathbf{k}',t_4) & = & \frac{1}{4} \biggr[ 4n(\mathbf{k},t_2)^2 + |m(\mathbf{k},\mathbf{k}',t_2)|^2 + |m(\mathbf{k}-2\mathbf{k}_L,\mathbf{k}'+2\mathbf{k}_L,t_2)|^2 \notag \\
 & & - m(\mathbf{k}-2\mathbf{k}_L,\mathbf{k}'+2\mathbf{k}_L,t_2)^*m(\mathbf{k},\mathbf{k}',t_2)e^{-i(\phi_L - \phi_R) - i\frac{\hbar}{2m}\left(|\mathbf{k}|^2 + |\mathbf{k}'|^2 - |\mathbf{k} - 2\mathbf{k}_L|^2 - |\mathbf{k}' + 2\mathbf{k}_L|^2 \right)\Delta t_{\mathrm{free}}} \notag \\
 & & - m(\mathbf{k},\mathbf{k}',t_2)^*m(\mathbf{k}-2\mathbf{k}_L,\mathbf{k}'+2\mathbf{k}_L,t_2)e^{i(\phi_L - \phi_R) + i\frac{\hbar}{2m}\left(|\mathbf{k}|^2 + |\mathbf{k}'|^2 - |\mathbf{k} - 2\mathbf{k}_L|^2 - |\mathbf{k}' + 2\mathbf{k}_L|^2 \right)\Delta t_{\mathrm{free}}} \biggr] . \label{eqn:C_D1D2_2}
\end{eqnarray}
\end{widetext}
where $\mathbf{k} \in \mathcal{V}(\mathbf{k}_1)$ and $\mathbf{k}' \in \mathcal{V}(\mathbf{k}_2)$ and $\Delta t_{\mathrm{free}} \equiv t_3 - t_2$ is defined as the duration of free-propagation 
between the $\pi$ and $\pi/2$ Bragg pulses. Having invoked Wick's theorem in Eq. (\ref{eqn:C_D1D2_2}), we may recognize that assuming 
the correlation function $G^{(2)}(\mathbf{k},\mathbf{k}',t_1)$ is a Gaussian function translates to the assumption that we may model the anomalous moment as
\begin{eqnarray}
 m\left( \mathbf{k},\mathbf{k}',t_2 \right) \equiv \bar{n}\sqrt{h} e^{i\theta(\mathbf{k},\mathbf{k}',t_2)} \prod_{d} e^{-(k_d+k_d')^2/4\sigma^2_d}, \label{eqn:m_model}
\end{eqnarray}
where the density of scattered atoms is assumed to be approximately homogeneous across the integration volumes and is given by the average $\bar{n}$. The argument $\theta(\mathbf{k},\mathbf{k}',t_2)$ of the complex 
anomalous moment is dependent on the specific model chosen for the collision, which we will elaborate upon momentarily.

Substituting Eq. (\ref{eqn:m_model}) into Eq. (\ref{eqn:C_D1D2_2}) gives the more recognizable form
\begin{eqnarray}
  G^{(2)}(\mathbf{k},\mathbf{k}',t_2) & = & \bar{n}^2 + \frac{\bar{n}^2h}{2}\prod_d \mathrm{exp} [-(k_d + k'_d)^2/2\sigma^2_d] \notag \\
 & & \times \left\{ 1 - \mathrm{cos}\left[ \phi_L - \phi_R + \varphi(\mathbf{k},\mathbf{k}') \right] \right\}, \;\; \;\; \; \label{eqn:G2_12}
\end{eqnarray}
where 
\begin{eqnarray}
 \varphi(\mathbf{k},\mathbf{k}') & = & \theta\left( \mathbf{k}-2\mathbf{k}_L, \mathbf{k}'+2\mathbf{k}_L, t_2 \right) - \theta\left( \mathbf{k}, \mathbf{k}', t_2 \right) \notag \\
 & & + \frac{\hbar}{2m} \left( |\mathbf{k}|^2 + |\mathbf{k}'|^2 - |\mathbf{k} - 2\mathbf{k}_L|^2 \right.  \notag \\
 & & - \left. |\mathbf{k}' + 2\mathbf{k}_L|^2 \right) \Delta t_{\mathrm{free}} . \label{eqn:varphi}
\end{eqnarray}
In comparison to the simple toy model of Eq. (\ref{eqn:G2_12_toy}) the most important new feature of Eq. (\ref{eqn:G2_12}) is the addition of $\varphi(\mathbf{k},\mathbf{k}')$, which acts 
as a momentum-dependent drift in the phase settings $\phi_L$ and $\phi_R$. As the phase settings are chosen to maximize the CHSH-Bell parameter, this new term can thus lead to a reduction in $S$. Composed of 
a free-propagation component and a dependence on the argument of the anomalous moment such an effect is similar to the phase dispersion of two-color photons in an earlier optical experiment of Rarity and Tapster \cite{Rarity-90}.

To investigate the impact of this new term and to evaluate the integral in Eq. (\ref{eqn:NiNj}) one must know the form of $\varphi(\mathbf{k},\mathbf{k}')$, which in turn 
explicitly depends on the argument $\theta(\mathbf{k},\mathbf{k}',t_2)$ of the anomalous moment. In general, this is not trivial as it requires an analytic solution of the anomalous moment from an appropriate model 
for the collision. To this end, we supplement our simple Gaussian-fit model by utilizing a solution of the anomalous moment based on a perturbative approach, previously used with success in 
Ref. \cite{Trippenbach:2008} (albeit for a different collision geometry---the BECs were split along the $x$ axis). Similar to the numerical treatment, this model takes into account the evolution 
of the spatial overlap of the split condensate wave-packets; however, it does not account for the spatial expansion of the condensates once released from the initial trap. 

To give a tractable form of the anomalous moment we approximate the initial mean field of the unsplit condensate as a Gaussian $\psi_0(\mathbf{x}) = \sqrt{\rho_0} \prod_d e^{-x^2_d/2\sigma^2_{g,d}}$ with 
peak density $\rho_0$ and rms widths $\sigma_{g,d}$ for $d=x,y,z$. The calculation of the anomalous moment is then straightforward and involves treating the wave-function of the scattered atoms 
with a perturbative expansion to low order. For a full derivation of the model we refer the reader to Ref. \cite{Trippenbach:2008}. In our solution we may make the approximation 
that the box sizes are sufficiently small such that $|\mathbf{k} - \mathbf{k}_1| \ll |\mathbf{k}_0|$ and $|\mathbf{k}' - \mathbf{k}_2| \ll |\mathbf{k}_0|$ and assume the condensates are completely 
spatially separated before applying the $\pi$ pulse, corresponding to $t_2/\tau_s \gg 1$ where $\tau_s = m\sigma_{g,z}/\hbar|\mathbf{k}_0|$ is the time-scale of separation. Under these limits the argument 
of the anomalous moment may be written as 
\begin{eqnarray}
 & & \theta\left( \mathbf{k}, \mathbf{k}', t_2 \right) \simeq -\frac{\hbar}{2m}\left( |\mathbf{k}|^2 + |\mathbf{k}'|^2 \right)t_2 \notag \\
 & & + \frac{\sigma_{g,z}}{\sqrt{\pi}|\mathbf{k}_0|} \left( \frac{|\mathbf{k}|^2 + |\mathbf{k}'|^2}{2} - |\mathbf{k}_0|^2 \right),
\end{eqnarray}
which thus allows us to write the phase drift as 
\begin{eqnarray}
\varphi(\mathbf{k},\mathbf{k}') & = & \left[ 8|\mathbf{k}_L|^2 - 4\mathbf{k}_L \cdot \left( \mathbf{k} - \mathbf{k}' \right) \right] \notag \\
& \times & \left[ \frac{\hbar}{2m}\left( \Delta t_{\mathrm{free}} - t_2 \right) + \frac{\sigma_{g,z}}{2|\mathbf{k}_0|\sqrt{\pi}} \right]. \label{eqn:phase_drift}
\end{eqnarray}

Using the form of Eq. (\ref{eqn:phase_drift}) and noting that our Bragg pulses couple only along the $k_y$ axis it is straightforward to evaluate the integral of Eq. (\ref{eqn:NiNj}),
\begin{eqnarray}
 C_{12} & = & \bar{n}^2 \prod_d \left(\Delta k_d\right)^2 + \frac{\bar{n}^2 h}{2} \prod_d \sigma_d\alpha_d \notag \\
 & &  - \frac{\bar{n}^2 h}{2}\left(\prod_d \sigma_d\right) \alpha_x\alpha_z\beta_y \mathrm{cos} \left(\phi_L - \phi_R \right) , \label{eqn:C_12_solved}
\end{eqnarray}
where $\alpha_d \equiv (e^{-2\lambda^2_d}-1) + \sqrt{2\pi}\lambda_d\mathrm{erf}(\sqrt{2}\lambda_d)$, $\lambda_d \equiv \Delta k_d / 2\sigma_d$, and 
\begin{eqnarray}
 \beta_y & \equiv & i\sqrt{\frac{\pi}{2}}\frac{e^{-8A^2|\mathbf{k}_L|^2\sigma^2_{y}}}{4A|\mathbf{k}_L|} \notag \\ 
 & & \times \Bigg[ e^{-i4A|\mathbf{k}_L|\Delta{k}_y} \mathrm{erf}\left( \frac{\Delta{k}_y + i4A|\mathbf{k}_L|\sigma^2_{y} }{\sqrt{2}\sigma_y} \right) \notag \\
 & & - e^{i4A|\mathbf{k}_L|\Delta{k}_y} \mathrm{erf}\left( \frac{\Delta{k}_y - i4A|\mathbf{k}_L|\sigma^2_{y} }{\sqrt{2}\sigma_y} \right) \notag \\
 & & + 2\mathrm{cos}\left(4A|\mathbf{k}_L|\Delta{k}_y\right)\mathrm{erf}\left(i2\sqrt{2}A|\mathbf{k}_L|\sigma_{y}\right) \Bigg], \label{eqn:beta_y} 
\end{eqnarray}
with $A \equiv \hbar(\Delta t_{\mathrm{free}} - t_{2})/2m + \sigma_{g,z}/2k_0\sqrt{\pi}$. 
One can then calculate the remaining correlation functions $C_{ij}$ in a similar fashion to find the correlation coefficient
\begin{eqnarray}
 E(\phi_L,\phi_R) \!& = &\! \left. \frac{C_{14} + C_{23} - C_{12} - C_{34}}{C_{14} + C_{23} + C_{12} + C_{34}} \right|_{(\phi_L,\phi_R)} \notag \\
 \!& = &\! \frac{h \alpha_x\beta_y\alpha_z}{ h\prod_d\alpha_d + 2 \prod_d \left(\lambda_d \right)^2 } \mathrm{cos}\left( \phi_L - \phi_R \right)\!. \;\; \;\;\;\; \label{eqn:E_solved}
\end{eqnarray}

The CHSH-Bell parameter is finally given by
\begin{eqnarray}
 S & = & 2\sqrt{2} \left|\frac{h \alpha_x\beta_y\alpha_z}{ h\prod_d\alpha_d + 2 \prod_d \left(\lambda_d \right)^2 } \right|. \label{eqn:S_solved}
\end{eqnarray}

\begin{figure}[tbp]
\includegraphics[width=8.0cm]{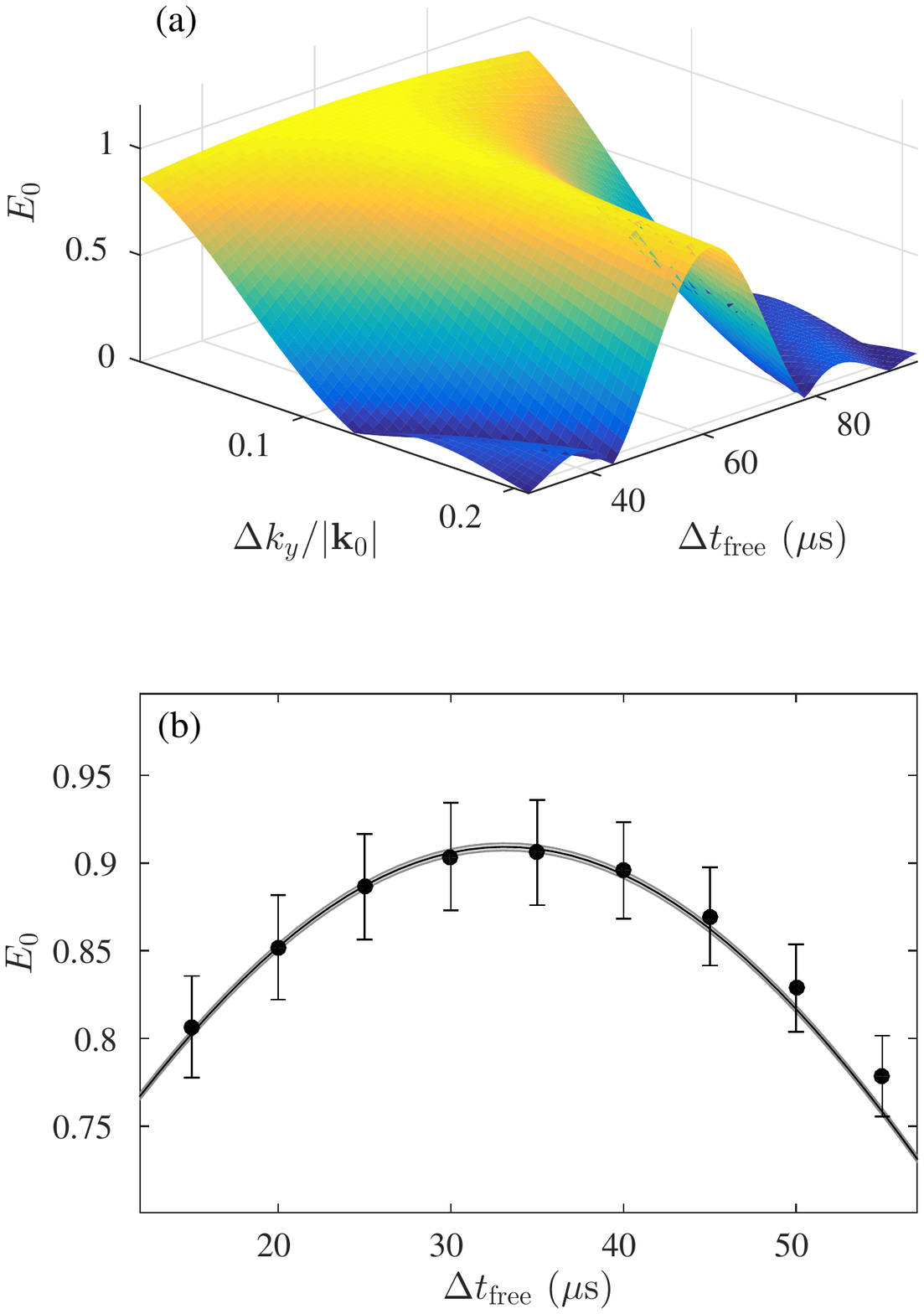}
\caption{(Color online) (a) Correlation amplitude $E_0$
predicted by the Gaussian-fit model [Eq. (\ref{eqn:S_solved})] as a function of the integration bin size $\Delta k_y$ and the free propagation time $\Delta t_{\mathrm{free}}$.
Calculations were performed for an initial condensate of $N = 1.9 \times 10^4$ atoms and other parameters as per the main text with $h$ and $\sigma_d$ extracted from the stochastic numerical results. 
The central ridge corresponds to Eq. (\ref{eqn:S_solved_t_opt}) where the phase drift term $\varphi(\mathbf{k},\mathbf{k}')$ is eliminated.
(b) Amplitude of the correlation function $E_0$ as a function of free propagation time $\Delta t_{\mathrm{free}}$ for an integration volume 
$(\Delta k_x, \Delta k_y, \Delta k_z) = (0.052,0.53,0.47)~\mu\mathrm{m}^{-1}$ and simulation parameters are as per (a). The predictions of the Gaussian-fit analytic model Eq.~(\ref{eqn:E_solved}) 
(gray shaded region)  are compared to the numerical results from stochastic simulations (black circles). The error bars on data points indicate the stochastic sampling error 
of two standard deviations obtained from $\sim \!800$ trajectories, while for the analytic prediction the uncertainty in $E_0$ (shaded region) 
is due to the uncertainty in the values $h$ and $\sigma_d$ extracted from the numerical simulations. 
} 
\label{fig:E_surf} 
\end{figure}

An important result of this model is the prediction that there exists an optimal free-propagation duration between the $\pi$ and $\pi/2$ Bragg pulses,
\begin{eqnarray}
 \Delta t_{\mathrm{free}} & = & t_2 - \frac{m\sigma_{g,z}}{\hbar k_0 \sqrt{\pi}}, \label{eqn:t_opt}
\end{eqnarray}
for which $\varphi(\mathbf{k},\mathbf{k}') = 0$ in Eq. (\ref{eqn:varphi}) for all $\mathbf{k} \in \mathcal{V}(\mathbf{k}_1)$ and 
$\mathbf{k}' \in \mathcal{V}(\mathbf{k}_2)$ and thus the phase settings retain their original values throughout the integration bin. This corresponds to 
$A = 0$ in Eq. (\ref{eqn:beta_y}) and we then find $\beta_y = \alpha_y$. Equation (\ref{eqn:S_solved}) is maximized under this condition and it transforms to
\begin{eqnarray}
 S & = & 2\sqrt{2} \frac{h \prod_d \alpha_d}{h\prod_d \alpha_d + 2 \prod_d \left(\lambda_d \right)^2}, \label{eqn:S_solved_t_opt}
\end{eqnarray}
where the dependence on box size is now characterised completely by the relative quantity $\lambda_d = \Delta k_d / 2\sigma_d$ for all directions, rather than the absolute length scale 
$\Delta k_y$ as in Eq. (\ref{eqn:S_solved}) along the $y$ axis.

\begin{figure}[tbp]
\includegraphics[width=8.0cm]{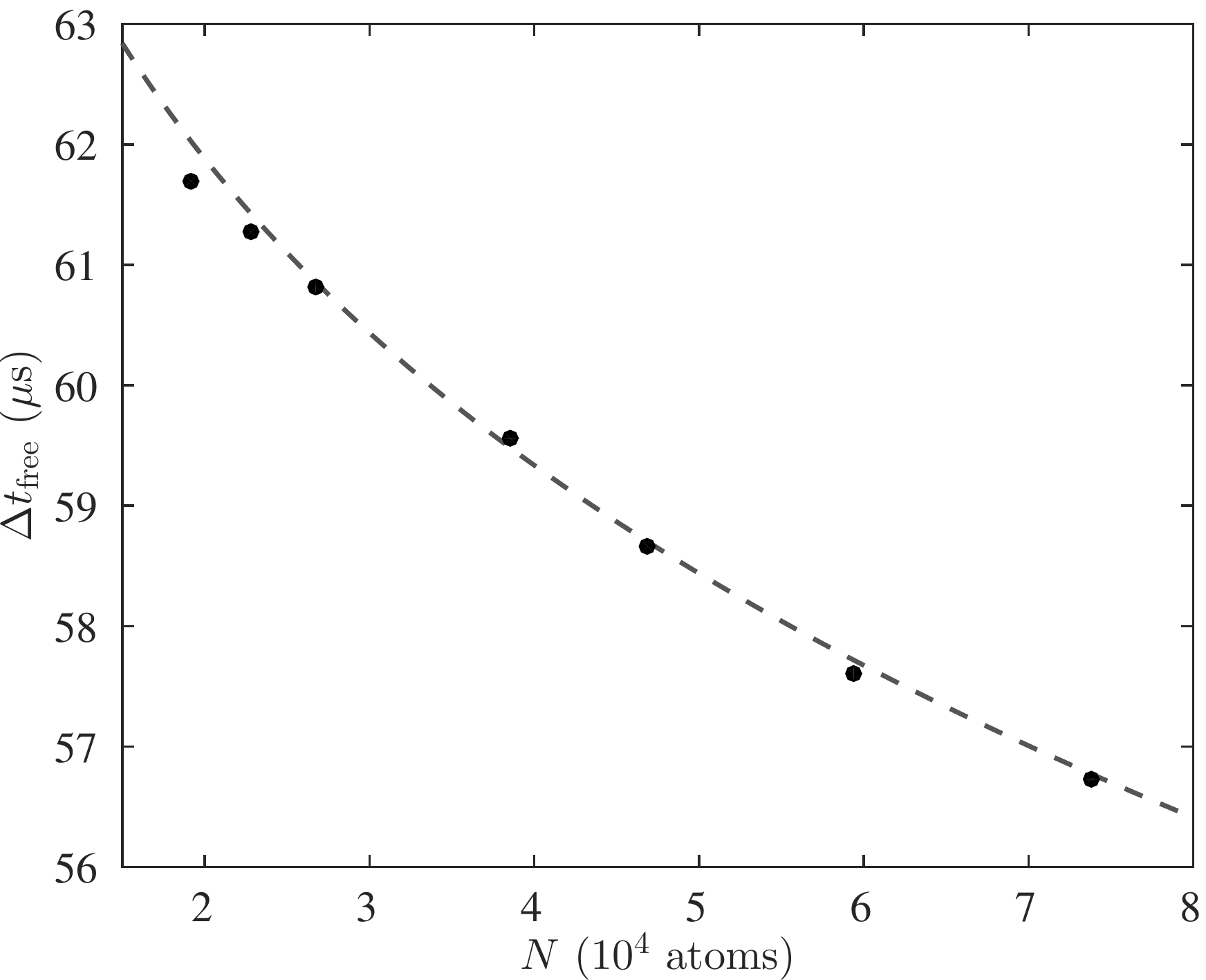}
\caption{Optimal free propagation time $\Delta t_{\mathrm{free}}$ for a range of initial BEC atom number.
Numerical results (black circles) are compared to the prediction of Eq. (\ref{eqn:t_opt}) from the perturbative model (dashed line). The range of $N$ in the initial BECs corresponds 
to those in the main text, while the integration volume is the same as Fig. \ref{fig:E_surf} (b).
} 
\label{fig:E_t_opt} 
\end{figure}

In Fig. \ref{fig:E_surf} (a) we plot Eq. (\ref{eqn:S_solved}) as a function of $\Delta{t}_{\mathrm{free}}$ and $\Delta k_y$ for the case of an initial BEC of $N = 1.9 \times 10^4$ atoms 
to illustrate the effects of the phase drift. As inputs to the model, the correlation height $h$ and correlation widths $\sigma_d$ are extracted from the numerical data at $t_1$, while the 
rms width $\sigma_{g,z}$ is chosen by fitting the numerically calculated trapped condensate to a Gaussian. For $\Delta{t}_{\mathrm{free}}$ satisfying Eq. (\ref{eqn:t_opt}), $S$ retains the maximal 
violation of Eq. (\ref{eqn:S_solved_t_opt}) with the strength only declining due to a dilution of the correlation as the integration box size $\Delta k_y$ increases. However, for $\Delta t_{\mathrm{free}}$ away 
from the optimal value one sees that an increase in the box size leads to a rapid decrease in $S$ due to rapid drift of the phase settings rather than a dilution of correlation. One can 
see this by noting that large $\Delta k_y$ implies the term $8|\mathbf{k}_L|^2 - 4|\mathbf{k}_L|\cdot(\mathbf{k}-\mathbf{k}')$ in Eq. (\ref{eqn:varphi}) will take large values near the edge of 
the integration volume and  $\varphi(\mathbf{k},\mathbf{k}')$ is scaled by this factor, leading to large deviations from the optimal phase settings. This is important as it demonstrates that for 
poor experimental resolution even small perturbations away from the optimal $\Delta{t}_{\mathrm{free}}$ can lead to a quick loss of Bell violation. 

Figure \ref{fig:E_surf} (b) shows results of stochastic numerical simulations for the amplitude of the correlation function $E_0$, where $E(\phi_L,\phi_R) \equiv E_0 \mathrm{cos} (\phi_L - \phi_R)$, as a 
function of $\Delta t_{\mathrm{free}}$ for the same initial BEC. We compare these results to the predictions of Eq. (\ref{eqn:E_solved}) to investigate the applicability of the Gaussian-fit model to a realistic system. 
We find excellent agreement, not only for the maximum attained correlation strength but also for the predicted optimal $\Delta t_{\mathrm{free}}$. The quantitative match to theory also implies that 
the underlying model for $\varphi(\mathbf{k},\mathbf{k}')$ is a good approximation to the form in the numerical simulations, although this is expected to break down for larger integration volumes where 
the assumptions for $\varphi(\mathbf{k},\mathbf{k}')$ in Eq. (\ref{eqn:phase_drift}) are no longer satisfied.

As the chosen phase angles $\phi_L$ and $\phi_R$ are shown to be unaffected in the final form of $E$ in Eq. (\ref{eqn:E_solved}), it is sufficient to numerically optimize 
$E_0$ as a function of $\Delta t_{\mathrm{free}}$ to maximize the Bell violation. In Fig. \ref{fig:E_t_opt} we plot the optimal $\Delta t_{\mathrm{free}}$ for a variety of initial BEC atom numbers determined 
from numerical calculations and compare these to the prediction of Eq. (\ref{eqn:t_opt}). Once again we find good quantitative agreement between the numeric and analytic methods. 
The numerically determined optimal $\Delta t_{\mathrm{free}}$ here are used in the simulations of the main text to define the timing of the application of the $\pi/2$ pulse.


%

\end{document}